\documentclass[prd,preprint,tightenlines,amsfonts,superscriptaddress]{revtex4}
\begin{document}

\title{Scaling behaviour in warm inflation}
\author{José P. Mimoso}
\affiliation{Departamento de F\'{\i}sica, Faculdade de Ci\^encias and Centro de F\'{\i}sica 
Te\'orica e Computacional, Universidade de Lisboa, Portugal}
\author{Ana Nunes}
\affiliation{Departamento de F\'{\i}sica, Faculdade de Ci\^encias and Centro de F\'{\i}sica 
Te\'orica e Computacional, Universidade de Lisboa, Portugal}
\author{Diego Pav\'{o}n}
\affiliation{Departamento de F\'{\i}sica, Universidad Aut\'onoma de Barcelona, Facultad de Ciencias, Bellaterra, Spain}
 
\begin{abstract}
We analyze the dynamics of warm inflation models with general viscous effects. We characterise the situations yielding asymptotic scaling behaviour.
\end{abstract}

\maketitle
\section{Introduction}
Scenarios leading the universe to a moderate temperature state at the end of the inflation avoiding the need of a separate stage of reheating~\cite{K+T 90,Reheat} have been proposed by several authors. It was advocated that this can be accomplished by coupling the inflaton to the radiation field in such a way that the decrease in the energy density of the latter during inflation is somewhat compensated by the decay of the inflaton into radiation~\cite{Berera+Ramos 03}. This kind of mechanism, first proposed by Berera~\cite{Berera 95}, was termed ``warm inflation" since the radiation temperature never drops dramatically. In addition, it has two further advantages: (i) the slow-roll condition $\dot{\phi}^2 \ll V(\phi)$ can  be fulfilled for steeper potentials, and (ii) the density perturbations generated  by thermal fluctuations may be larger than those of quantum origin~\cite{origin}. 

One natural further step is to assume that the inflaton field decays not just into radiation but into massive particles as well. The mixture of massless and massive particles can be described by an overall fluid  with equation of state $p = (\gamma- 1)\rho$, where the adiabatic index $\gamma$ is bounded by $1 \leq \gamma \leq 2$. It is also natural to expect that this fluid has a negative dissipative pressure, $\Pi$, that quantifies the departure from thermodynamical equilibrium. This viscous pressure arises on rather general grounds from interparticle interactions~\cite{radiative}. 

Our purpose in this work is to analyse the dynamics of models that generalize the usual warm inflationary scenario by introducing these novel elements, namely  the decay of the scalar field into a fluid of adiabatic index $\gamma$ rather than just radiation, and specially the dissipative pressure of this fluid. We will not dwelve into the difficult question of the quantum, nonequilibrium thermodynamical problem underlying warm inflation~\cite{Quantum therm}, but rather take a phenomenological approach similar to that considered in several works~\cite{phenomenology,Billyard+Coley 00}. We allow for various forms of the rate of decay of the scalar field, as well as for various expressions of the dissipative pressure, and resort to  the qualitative analysis of the corresponding autonomous system of differential equations using the approach developed in~\cite{NM00}. This enables us to consider arbitrary scalar field potentials. 

A central question we address is the existence of scaling solutions~\cite{Billyard+Coley 00,
NM00,scaling}, i.e., solutions where the ratio of energies involving the matter fluid and scalar field keep a constant ratio. In the standard treatment of warm inflation the existence of a ``quasi-static'' regime for the energy transfer between the scalar field and the radiation is assumed for the computation of the temperature of matter at the final stages of warm inflation~\cite{origin}. It is however unclear why there should be such a quasi-static stage in the conventional model of warm inflation, and thus whether such an assumption is fully legitimate. If there were to be a scaling regime at the end of warm inflation such a quasi-static regime would necessarily exist. Moreover, this class of solutions  finds itself a justification in the kinetic analysis of interacting fluids~\cite{Thermodynamics}. A particular illustration of this can be found in~\cite{Lima+Carrillo}.

In this work we show that the existence of a late time scaling behaviour depends not only on the asymptotic form of the inflaton potential~\cite{NM 00}, but also on having a time-dependent rate of decay for the scalar field. 

\section{The dynamical system}
We consider a  spatially flat Friedmann-Robertson-Walker Universe filled with a self-interacting scalar field and a perfect fluid consisting in a mixture of matter and radiation, such that the former decays into the latter at some rate $\Gamma$ and that the latter exhibits a dissipative pressure $\Pi$. On the other hand, we neglect radiative corrections to the inflaton potential. The  corresponding system of equations reads
\begin{eqnarray}
3H^2 &=& \rho + \frac{\dot{\phi}^2}{2} + V(\phi)\label{eq_Fried}\\     
\dot{H}&=& - \frac{1}{2}\,\left(\dot{\phi}^2+\gamma \rho+\Pi\right) \label{eq_Ray}\\
\ddot{\phi} &=& - (3H+\Gamma)\dot{\phi}-V'(\phi)  \label{eq_KG}\\
\dot{\rho} &=& -3\,H\, (\gamma\rho + \Pi)+ \Gamma \dot{\phi}^2  \label{eq_cons}\; ,
\end{eqnarray}
where we have used units such that $8\pi G= c^2=1$. The first two are Einstein's equations, the third  describes the decay of the inflaton, and the fourth one is the energy balance for the matter fluid. As usual $H \equiv \dot{a}/{a}$ denotes the Hubble factor.

To cast the corresponding autonomous system of differential equations it is expedient to introduce the set of normalized, dimensionless variables
\begin{equation}
x^2 = \frac{\dot{\phi}^2}{6H^2}\quad , \qquad
y^2 = \frac{V(\phi)}{3H^2}\quad , \qquad
r = \frac{\Gamma}{3H} \; ,
\end{equation}
as well as $\chi= \Pi/3H^2$, along with the new time variable $N= \ln{a}$. 
We assume, quite generally, $\Gamma_\phi= \tilde\Gamma(\phi)\,H^\delta$ and $\Pi = -3\zeta\,\rho^\alpha\, H^\beta$, where $\delta$, $\zeta$, $\alpha$ and $\beta$ are constants and moreover $2\alpha+\beta-2=0$ on dimensional grounds. The latter condition on the parameters $\alpha$ and $\beta$ implies that the dimensionless variable $\chi$ becomes $\chi=-3^{\alpha}\zeta\, (\rho/3H^2)^\alpha$, and hence reduces to $\chi=-3^{\alpha}\zeta\, (1-x^2-y^2)^\alpha$, since $\rho/3H^2=1-x^2 - y^2$. Therefore the dynamical system is 4-dimensional and reads
\begin{eqnarray}
x' &=& x \,\left( Q-3 (1+r)\right)- W(\phi) \, y^2, \label{x'_5} \\  
y' &=& \left(Q+W(\phi)\, x\right) y\; , \label{y'_5}\\
r'  &=& r \left[ \sqrt{6}\left(\frac{\partial_\phi\tilde\Gamma}{\tilde\Gamma}\right)\, x + Q (1-\delta) \right] \label{r'_5}\\
\phi' &=& \sqrt{6} \, x \; , \label{phi'_1}
\end{eqnarray}
where a prime indicates differetentiation with respect to $N$, and where we have used the definitions $
 W(\phi) =\sqrt{\frac{3}{2}}\, \left(\frac{\partial_{\phi}V}{V}\right)$
 and 
\begin{equation}
Q =\frac{3}{2}\,\left[2x^2+\gamma\,
(1-x^2-y^2)- 3^{\alpha}\zeta\, (1-x^2-y^2)^\alpha\right] \label{defQ_1} \; .
\end{equation}
Equation (\ref{phi'_1}) was first considered in \cite{NM00} and is crucial for the consideration of general potentials $V(\phi)$ besides the particular case of the exponential potential. The function $Q$ defined by  Eq.~(\ref{defQ_1}) is related to the deceleration parameter $q=-\ddot{a}a/\dot{a}^2$ by $q=Q-1$. 

\section{Qualitative analysis}
\subsection{The special case $\Gamma_\phi \propto H$ and $\Pi=0$}\label{Particular case}
Here we take $\Gamma_\phi =3r H$, where $r$ is a positive constant. The dynamical system reduces to the three equations
\begin{eqnarray}
x' &=& x \,\left[Q-3(1+r)\right]- W(\phi) \, y^2, \label{x'_2} \\  
y' &=& \left[Q+W(\phi)\, x\right] y\; , \label{y'_2}\\
\phi' &=& \sqrt{6} \, x \; , \label{vphi'_2}
\end{eqnarray}
where $Q$ is now simply given by $Q =\frac{3}{2}\,\left[2x^2+\gamma\,
(1-x^2-y^2)\right]$. 

We see that these equations are analogous to those of the $r=0$ case with a different coefficient on the linear term in $x$ of Eq.~(\ref{x'_2}).
Thus the basic qualitative dynamical features remain the same as those found for that model~\cite{Billyard+Coley 00,NM00,scaling}. The decay of the scalar field introduces though two major consequences. 

The origin $x=0$, $y=0$ is again a fixed point associated with the vanishing of the scalar field's energy and, hence, corresponding to matter domination, but the interaction given by a non-vanishing $r$ has the relevant effect (already found in the constant $\Gamma$ model) that the stability of the minima is reinforced and that the maxima become less unstable. 
The other major effect of the interaction arises when we look for fixed points at $\phi \to \infty$. Now, we find that there are always attracting scaling solutions for potentials that have an asymptotic exponential behaviour, that is, for potentials for which $W\to {\rm const}$ when $\phi \to \infty$~\cite{NM00}. Moreover, this happens independently of the steepness of the late time exponential behaviour, a remarkable effect of the present model.

Assuming the potential to be asymptotically given by $V\propto e^{-\lambda \phi}$, these solutions are $a(t) \propto t^A$, $\phi-\phi_0 = \ln{t^{\pm 2/\lambda}}$, where $A$ is given in implicit form by
\begin{equation}
3\gamma\left(A-\frac{2}{3\gamma}\right)\left( A-\frac{2}{\lambda^2}(1+r)\right) - \frac{4r}{\lambda^2}=0 \; .
\end{equation}
A linear expansion in $r$ in the neighborhood of the scaling solutions of the $r=0$ case yields 
\begin{equation}
A=\frac{2}{3\gamma}\,\left[ 1+\frac{\frac{2}{\lambda^2}}{\left(\frac{2}{3\gamma}-\frac{2}{\lambda^2}\right)}\,r\right] \; .
\end{equation}
So the decays have the effect of increasing (resp. decreasing) the scale factor rate of expansion with regard to the $r=0$ case if $\lambda^2>3\gamma$ (resp. $\lambda^2<3\gamma$). In particular  we can see that the scaling behaviour can be inflationary, for cases where this would not happen in the absence of decays. For instance, taking $\gamma=4/3$ and $\lambda^2 >4$ the condition for the scaling solution to be inflationary is $1+r>\lambda^2/4 >1$. Thus, in this model, the solutions yield endless power-law inflation even for a modest scalar field decay, provided that the asymptotic behaviour of the potential is steep enough, i.e., $\lambda^2>3\gamma$ ($>4$ in the present example). Besides the scaling solutions,  there are also fixed points corresponding to de Sitter behaviour $x=0$, $y=1$, whenever the scalar field potential exhibits an asymptotic, non-vanishing constant value. However, when the potential is asymptotically exponential, there are no fixed points on the boundary $x^2+y^2=1$ at $\phi \to \infty$, labelled $\phi_\infty$, in contrast to what happens in the $r=0$ case.

From a thermodynamical viewpoint the former scaling solutions are particularly interesting. In a universe with two components it can be shown~\cite{Thermodynamics} that the temperature of each of the components satisfies the equation
\begin{equation}
\frac{\dot{T}_{i}}{T_{i}}= -3\frac{\dot{a}}{a} \left(
1-\frac{\Gamma_{i}}{3H}\right) \frac{\partial p_{i}}{\partial \rho_{i}}+\frac{n_i \dot{s}_i}{\partial \rho_i/\partial T_i}
\; ,
\end{equation}
where ($i=1,2$), $n_i$ denotes the number density of particles of the $i$-species,  $\Gamma_i$ their rate of change, and $T_i$ the temperature of this component. In the important case of particle production happens with constant entropy per particle, $\dot s_i=0$, we also have $(\rho_1+p_1)\Gamma_1=-(\rho_2+p_2)\Gamma_2$.
Thus, taking the first component to be the matter/radiation fluid, and the second to be the inflaton scalar field, we have 
\begin{equation}
\Gamma_\phi=\frac{(\rho+p)}{\dot{\phi}^2}\, \Gamma_{m/r} \propto \Gamma_{m/r}\; .
\end{equation}
As $(\rho+p)/\dot{\phi}^2=\gamma(1-x^2-y^2)/x^2$ is a constant at the scaling solutions, then $\Gamma_\phi =3rH$, with $r$ a constant,  implies $\Gamma_{m/r}
=3\sigma H$, where $\sigma$ is another constant that depends both on $r$ and on the location of the scaling solution. This yields a temperature of the matter/radiation  component evolving as a power-law $T\propto a^{-(\gamma-1)(1-\sigma)}$. Thus for $\sigma$ close to 1, the temperature of the matter/radiation remains quasi-static, whereas for $\sigma>1$ (resp. $\sigma<1$) it increases (resp. decreases)\footnote{Notice also that, for $\sigma\simeq 0$, we recover the temperature law for perfect fluids without dissipative effects.}. Provided that we guarantee enough inflation, $r$ need not be very large (contrary to what is usually assumed to facilitate  the slow-rolling). 

\subsection{The general case}\label{General case}
In what follows we return to the general system~(\ref{x'_5})-(\ref{phi'_1}). It seems reasonable to further assume that $0<\alpha<1$ so that $\beta>0$ which amounts to having a bulk viscosity pressure whose importance diminishes with the expansion and with the dilution of matter.

At finite values of $\phi$ we find a line of fixed points $x=y=r=0$ ($\phi$ is arbitrary), and a line of fixed points characterized by $x=0$, $\phi=\phi_0$, $(1-y^2)^{1-\alpha}=3^{\alpha}\zeta/\gamma$ and any value of $r$. In the latter case $\phi_0$ is the value of $\phi$ at an extremum of $V(\phi)$, i.e., where $W(\phi_0)=0$, and we require that 
\begin{equation}
3^{\alpha}\zeta <\gamma \label{gen_cond1}\; .
\end{equation}

The linear stability analysis shows that the singular points $x=y=r=0$ corresponding to matter domination are unstable. Regarding the  line of singular points with $r\neq 0$, linear stability analysis reveals that, besides the vanishing eigenvalue associated with $r$ ($\lambda_r=0$), the sability is once more determined by the nature of the extremum of $V(\phi)$. Indeed, one eigenvalue is always negative, and the eigenvalues along $x$ and $\phi$ are given by 
\begin{equation}
\lambda_{x,\phi} = - \frac{3(1+r)}{2}\pm \frac{1}{2}\,\sqrt{9(1+r)^2-4\sqrt{6}\,y_\ast^2\,W'(\phi_0)} \, , \label{eigenv_x5}
\end{equation}
where $y_\ast$ is the equilibrium $y$-coordinate.
We see from the latter equation that $\lambda_{x,\phi}>0$ requires
that $W'(\phi_\ast)<0$, that is a maximum at $V(\phi_\ast)$. Otherwise, in the case of a minimum of $V(\phi)$, we have either a stable node  (when $0<W'(\phi_0)<9(1+r)^2/4\sqrt{6}\,y_\ast^2$) or a stable sink (when $W'(\phi_0)>9(1+r)^2/4\sqrt{6}\,y_\ast^2>0$). 
Decreasing $y_\ast$ induces an effective reduction of the steepness of the potential at the maxima and a greater stability of the minima. The net effect is that the system spends more time in the neighborhood of a fixed point associated with a  maximum of the potential (alternatively, the minima become more stable). This is helpful  for setting the conditions for slow-roll inflation, which translate as  $3(1+r)>y_\ast$, and do not require a large rate of decay of the scalar field. 

At $\phi_\infty$, Eq.~(\ref{r'_5}) shows that we may have the usual fixed points corresponding to a non-vanishing, flat asymptotic asymptotic behaviour of the potential (late-time approach to a cosmological constant) if $Q=0$, $\Gamma'/\Gamma=0$ and $\Pi'/\Pi=0$ simultaneously. However, we also find asymptotic scaling behaviour in the case $W(\phi)$ approaches an exponential behaviour ($W(\phi_\infty)=-\lambda$, with constant $\lambda>0$), provided
\begin{equation}
W(\phi) = \frac{\sqrt{6}}{1-\delta}\,\left(\frac{\partial_\phi \tilde\Gamma}{\tilde\Gamma}\right)\; ,
\end{equation}
which amounts to having
$\displaystyle \tilde\Gamma(\phi_\infty)=\nu\, \left(V(\phi_\infty)\right)^{\frac{1-\delta}{2}}$, where $\nu$ is a constant. We thus effectively have a $\Gamma_\phi \propto H$ rate of decay, as in the previous subsection. Hence this emerges a necessary condition for scaling behaviour. In polar coordinates  $x=\bar{\rho}\, \cos{\theta}$, $y=\bar{\rho}\, \sin{\theta}$, these scaling solutions are given by
\begin{equation}
\cos{\theta_\ast} = \frac{\lambda}{3(1+r)}\,\sqrt{u}\; , \label{sp_theta_gen5}
\end{equation}
where $u\equiv \bar{\rho}^2$ is a root of the quadratic equation
\begin{equation}
(1-u)\,(a-b u) - r b u - \frac{3^{\alpha}}{2}\,(1-u)^\alpha =0 \; , \label{sp_rho_gen5}
\end{equation}
and $a = \frac{\gamma}{2}$ and $b= \frac{\lambda}{3(1+r)}$.

We see from Eqs.~(\ref{sp_theta_gen5}) and (\ref{sp_rho_gen5}) that there is always one (and only one) scaling solution, provided the condition~(\ref{gen_cond1}) holds. (Notice that this was precisely the condition that was required for the existence of fixed points at finite $\phi$). 
We also find that the location of the singular points corresponding to the scaling behaviour is now closer to $x^2+y^2=0$, having a smaller $y_\infty$ value than in the models without bulk viscosity. Moreover, linear stability analysis reveals that under the conditions (\ref{gen_cond1}) and $\alpha<1$ the scaling solutions are stable, i.e., they are attractors. These results mean that the late time contribution of the matter component is enhanced by the viscous pressure. This is a most convenient feature for the warm inflation scenario, since it further alleviates the depletion of matter during inflation and the subsequent need for reheating.

\acknowledgments{
J.P.M. and A.N. wish to acknowledge the finantial support from the Funda\c c\~ao para a 
Ci\^encia e a Tecnologia under the POCTI/FNU/49511/2002 grant.}

\end{document}